\documentstyle[12pt,psfig]{article}
\def\etal{et al.~}
\def\ang{\ifmmode {\rm \AA} \else $\rm \AA$\fi}  

\def\hii{H~{\sc ii}}
\def\hei{He~{\sc i}}
\def\heii{He~{\sc ii}}

\def\he0{\ifmmode {\rm He^{\circ}} \else $\rm He^{\circ}$\fi}
\def\hep{\ifmmode {\rm He^+} \else $\rm He^{+}$\fi}
\def\hepp{\ifmmode {\rm He^{2+}} \else $\rm He^{2+}$\fi}
\def\halpha{\ifmmode {\rm H{\alpha}} \else $\rm H{\alpha}$\fi}
\def\hbeta{\ifmmode {\rm H{\beta}} \else $\rm H{\beta}$\fi}
\def\hgamma{\ifmmode {\rm H{\gamma}} \else $\rm H{\gamma}$\fi}

\def\masl{\ifmmode  {\rm M_{\sun}yr^{-1}} \else ${\rm M_{\sun}yr^{-1}}$\fi}

\def\mdot{\ifmmode  \dot{M} \else $\dot{M}$\fi}
\def\msun{\ifmmode M_{\odot} \else $M_{\odot}$\fi}
\def\vinf{\ifmmode v_{\infty} \else $v_{\infty}$\fi}
\def\teff{\ifmmode T_{\rm eff} \else $T_{\rm eff}$\fi}
\def\logg{\ifmmode \log g \else $\log g$\fi}
\def\loggeff{\ifmmode \log g_{\rm eff} \else $\log g_{\rm eff}$\fi}
\def\rstar{\ifmmode R_{\star} \else $R_{\star}$\fi}
\def\lstar{\ifmmode L_{\star} \else $L_{\star}$\fi}
\def\mstar{\ifmmode M_{\star} \else $M_{\star}$\fi}
\def\rsun{\ifmmode R_{\odot} \else $R_{\odot}$\fi}
\def\lsun{\ifmmode L_{\odot} \else $L_{\odot}$\fi}
\def\zsun{\ifmmode Z_{\odot} \else $Z_{\odot}$\fi}
\def\12c16o{$^{12}{\rm C}\left(\alpha,\gamma\right)^{16}{\rm O}$}
\def\kms{\ifmmode {\rm km \;s^{-1}} \else $\rm km \;s^{-1}$\fi}
\def\nlte{non--LTE}
\title{\bf WR Populations in Starbursts: WN and WC Subtypes
	and the Role of Binaries}
\author{Daniel Schaerer $^1$ \\ and \\William D. Vacca $^2$\\
\vspace{1cm}\\
\normalsize $^1$Space Telescope Science Institute, Baltimore MD 21218, USA 
		(schaerer@stsci.edu) \\
\normalsize $^2$Institute for Astronomy, Honolulu, HI 96822, USA
		(vacca@athena.ifa.hawaii.edu)}
\date{\mbox{}}
\begin{document}
\maketitle
\pagestyle{empty}
%
%
\def\bull{\vrule height .9ex width .8ex depth -.1ex}
\makeatletter
\def\ps@plain{\let\@mkboth\gobbletwo
\def\@oddhead{}\def\@oddfoot{\hfil\tiny\bull\quad
``WR Stars in the Framework of Stellar Evolution'';
33$^{\mbox{\rm rd}}$ Li\`ege\ Int.\ Astroph.\ Coll., 1996\quad\bull}%
\def\@evenhead{}\let\@evenfoot\@oddfoot}
\makeatother
%
%
\def\beginrefer{\section*{References}%
\begin{quotation}\mbox{}\par}
\def\refer#1\par{{\setlength{\parindent}{-\leftmargin}\indent#1\par}}
\def\endrefer{\end{quotation}}
%
%
{\noindent\small{\bf Abstract:} 
We present the first results of a new set of population synthesis models, 
which utilize the latest stellar evolutionary tracks, recent non-LTE
atmosphere models which include stellar winds, and observed line strengths 
in WR spectra to predict the relative strengths of various WN and
WC/WO emission features in the spectra of starburst galaxies. 
Our results will be used to derive accurate numbers of WN and WC stars
in starburst galaxies.
We also analyze the frequency and the WN and WC content of WR-rich galaxies 
in low metallicity samples; the theoretical predictions are found to be in 
good agreement with the observed frequencies.
We also discuss the possible role of massive close binaries in starburst regions. 
If the starburst regions are formed in relatively instantaneous bursts
we argue that, given their young age as derived from emission lines equivalent widths, 
{\em (1)} in the majority of the observed WR galaxies massive close binaries
	have not contributed significantly to the WR population, and
{\em (2)} nebular \heii\ 4686 emission is very unlikely due to massive X-ray 
	binaries.
}
%
%
\section{Introduction}
The presence of large numbers of Wolf-Rayet (WR) stars in extragalactic star-forming 
objects (hereafter called WR-galaxies)
of quite heterogeneous types is well established (see e.g.~the compilation of 
Conti 1991). New serendipitous discoveries of WR galaxies have resulted from studies 
covering a wide range of topics, from the primordial He abundance determination 
(cf.~Izotov et al.\ 1996a) to the nature of Seyfert galaxies (Heckman et al.\ 1996), 
and a considerable number of new observations can be expected with the new generation 
of 8-10 m class telescopes.

In most cases the presence of WR stars can be used as a powerful constraint 
on the age of the starburst episode (typically $3 - 8$ Myr). The luminosity 
in the broad ``WR-bump'' (centered at $\lambda$ 4650) can be used to derive the 
total number of WR stars present in the burst. 
From the strength of the nebular emission lines one can also determine the
numbers of OB stars, which are the dominant contributors to the Lyman continuum flux.
Additional information on the slope of the Initial Mass Function (IMF) 
can also be obtained (Meynet 1995; Contini et al.~1995; Schaerer 1996).
When compared with evolutionary models, the derived WR/O number ratios indicate 
that star formation occurs in ``bursts'' short compared to the lifetime 
of massive stars (Arnault et al.~1989; Vacca \& Conti 1992; Meynet 1995). 
We refer the reader to the contribution by Vacca in these proceedings for a 
more detailed discussion of the properties and analysis of WR galaxies.

Many aspects of WR galaxies remain to be explored and there are several 
questions regarding the effect of large numbers of WR stars on their host
galaxies that remain unanswered. Among these are the following, which we hope to 
address in this contribution:
What fraction of starbursts have gone through or are currently in 
a WR-rich phase (Kunth \& Joubert 1985; Meynet 1995) ?
How frequent are WC stars in WR-galaxies and what is their importance
(Meynet 1995; Schaerer 1996) ? Are WR stars responsible for the high excitation 
nebular lines observed in the optical spectra of some young starbursts (Garnett et 
al.~1991; Motch et al.~1994; Schaerer 1996) ? How important is the formation of WR stars 
in binary systems for WR-galaxies (Cervi\~{n}o \& Mas-Hesse 1996; Vanbeveren et al.~1996) ?

Answers to these questions require a detailed knowledge of the stellar populations in 
the host galaxies. The aim of our work is to provide new predictions for the WR populations
in young starbursts, by explicitly taking into account the two main W-R subtypes, 
WN and WC stars. Our synthesis approach, based on well-tested evolutionary models, 
recent atmosphere models for O and WR stars, and observed line-strengths in WR stars,
provides a number of relevant observable quantities. (A similar, but less comprehensive
attempt, was carried out by Kr\"uger et al.\ 1992.)
Here we present preliminary results from this on-going work, which will be used
for the future analysis of a large sample of WR-galaxies. 
It is our hope to shed some light on some of the aforementioned questions.

\section{Evolutionary synthesis models}
In this Section we briefly describe the adopted model ingredients for
our evolutionary synthesis models, the most important input parameters,
and the synthesized quantities.

{\em Stellar evolution:} We adopt the recent tracks of the the Geneva
group, which cover the metallicity range from $Z$=0.001 (1/20 \zsun) to 
$Z$=0.04 (2 \zsun) (see Meynet \etal 1994 and references therein). 
As shown by Maeder \& Meynet (1994) the models with enhanced mass loss 
rates reproduce a large number of observations regarding massive star populations, 
including WR/O star ratios in various nearby galaxies.
These models are preferred over the earlier models of Schaller \etal 
(1992) adopted in the calculations of Vanbeveren (1995) and other 
population synthesis models (e.g.~Cervi\~{n}o \& Mas-Hesse 1994, 1996).

{\em Evolution of massive close binaries:}
In order to explore the effects of forming WR stars via
mass transfer in massive close binaries on the total population of massive stars,
we adopted the following simplified treatment of binaries.
We used the recent calculations for various metallicities of de Loore 
\& Vanbeveren (1994), who assume an initial mass ratio of 0.6, Case B 
mass transfer, and neglect the possibility of subsequent WR formation by 
the secondary.
For our synthesis models one free parameter $f$ determines the binary
population; $f$ is defined as the fraction of stars, which are primaries 
in close binary systems and which
will therefore experience Roche lobe overflow during their evolution.
The total WR population formed through the binary channel and the 
distribution among the different subtypes can be derived directly 
from the stellar lifetimes and the duration of the respective phases 
(see de Loore \& Vanbeveren 1994).
To determine the impact of binaries on observational properties (nebular 
lines and broad WR emission lines) we adopted an average Lyman continuum 
flux of $Q_0 = 10^{49} \, {\rm photons \, s^{-1}}$ which roughly corresponds 
to the average contribution 
from single WR stars at the time during the evolution of a burst population 
when binary stars are first expected to be formed.
The broad WR line emission is treated in the same manner as for single stars (see below).

{\em Continuum spectral energy distribution:}
To determine the stellar continuum spectral energy distribution at each time 
during a burst, we relied on three different sets of theoretical models:
{\em 1)}
For massive stars we used the spectra from the combined stellar structure and atmosphere
({\em CoStar}) models of Schaerer \etal (1996ab), which
include \nlte\ effects, line blanketing, and stellar winds. 
These models cover the entire parameter space of O stars during their main sequence
evolution.
{\em 2)}
For later spectral types we use the line-blanketed plane-parallel 
LTE models of Kurucz (1992).
{\em 3)}
For W-R stars, we used the spherically expanding
\nlte\ models of Schmutz, Leitherer \& Gruenwald (1992).
In addition to the stellar continuum one also needs to account for
the nebular continuous spectrum. Its emission is calculated assuming
$T_e=$ 10 kK, $N_e=100 \, {\rm cm^{-3}}$, and solar H/He abundances. 

{\em Nebular and WR emission lines:}
The strengths of the nebular recombination lines (primarily \hbeta, \halpha, 
\heii\ $\lambda$ 4686) are calculated with the same values of the electron 
temperature and density as used for the nebular continuum. We have compiled 
average {\em stellar} line fluxes of the 
strongest WR emission lines for WN, WC, and WO stars. 
We distinguish 5 WC subtypes as well as the WO subtype, as these objects show 
considerable differences in their line fluxes.
We also include possible emission from OfI stars.
The line fluxes have been taken from the following sources:
Crowther (1996, private communication) and Smith \etal (1996) 
for WN stars, and Smith \etal (1990ab) for WC/WO stars.
More details are given in Schaerer \& Vacca (1996).
The WR stage, including the WC subtype (see Smith \& Maeder 1991), 
is determined by the surface abundances predicted from the evolutionary models.

{\em Input parameters:}
In the present work we consider the time evolution of an instantaneous
burst of star-formation. The basic parameters of our models are therefore
the metallicity, the binary fraction, and the slope and upper mass cut-off of the
initial mass function. Here, we adopt a Salpeter IMF with
an upper mass cut-off of 120 \msun. The results do not depend
on the lower mass cut-off, as long as it is less than about $5$ \msun. 
Variations of the IMF slope are considered in Schaerer \& Vacca (1996).

{\em Synthesized quantities:}
The major predictions from our models include:
{\em (1)} the relative populations of O stars (where an O stars is defined by 
\teff\ $>$ 30 kK), WN stars, and WC/WO stars,
{\em (2)} emission line fluxes and equivalent widths of the following
broad WR lines: \heii\ $\lambda$ 1640, N~{\sc iii} $\lambda$ 4640,
C~{\sc iii/iv} $\lambda$ 4650, \heii\ $\lambda$ 4686, total $\lambda$
4650 WR-bump, C~{\sc iv} $\lambda$ 5696, and C~{\sc iv} $\lambda$ 5808,
{\em (3)} the WR contributions to \halpha\ and \hbeta,
{\em (4)} the ionizing photon fluxes in the H, \hei, and \heii\ continua, and
{\em (5)} the emission line fluxes and equivalent widths of nebular lines of
\halpha, \hbeta, and \heii\ $\lambda$ 4686.

\begin{figure}[htb]
\centerline{\psfig{figure=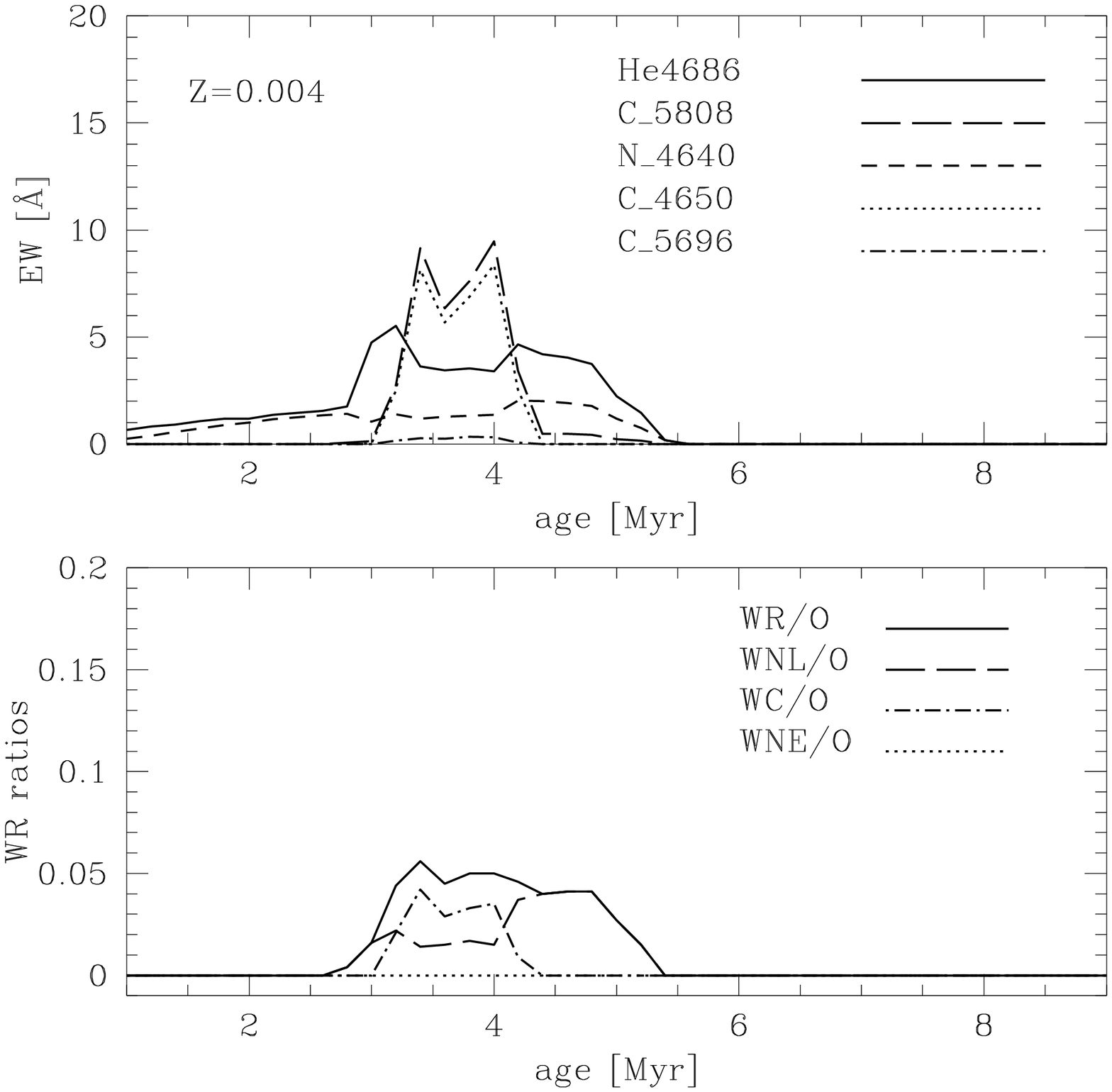,height=8cm}
	\psfig{figure=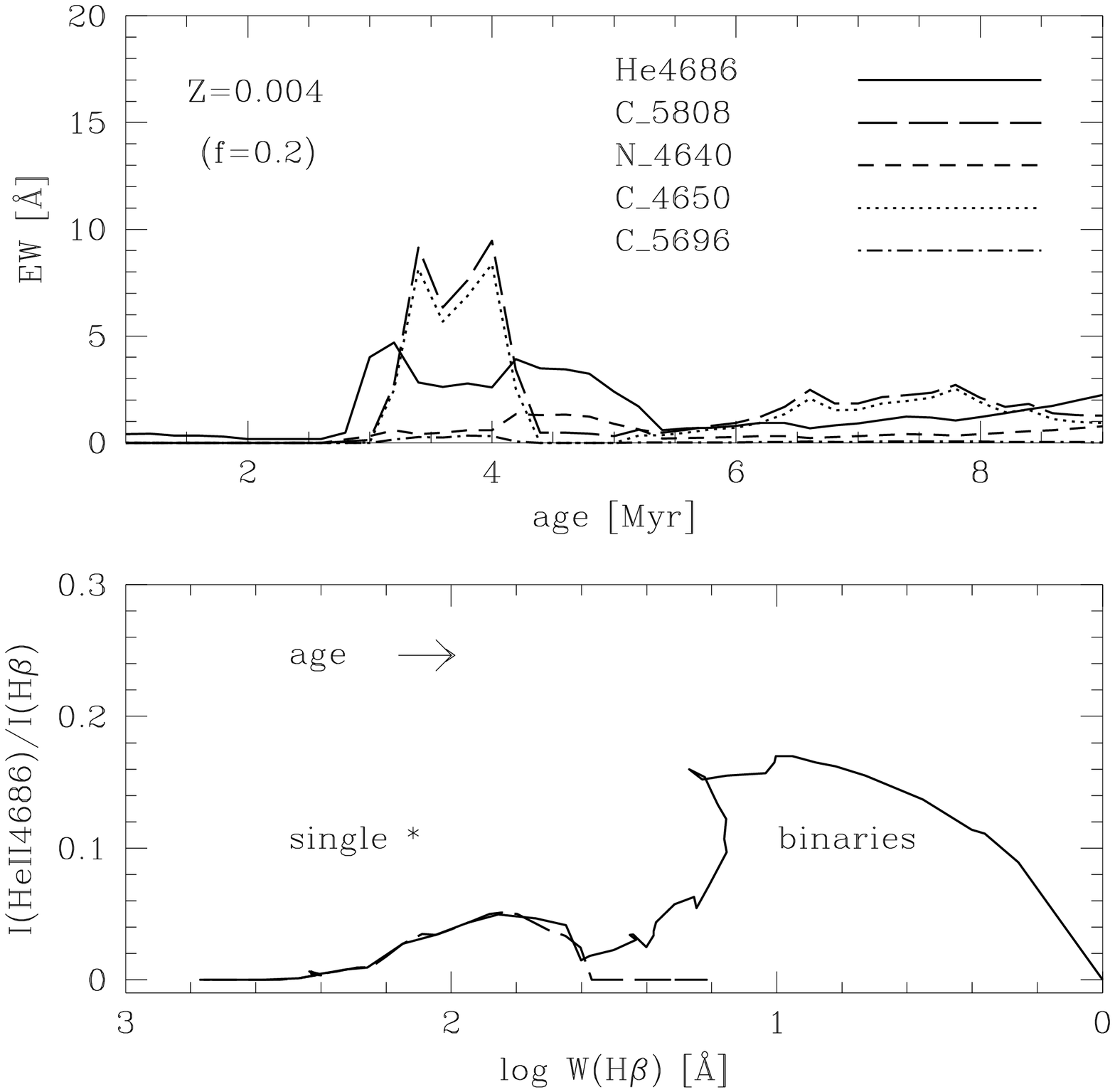,height=8cm}}
\caption{{\em Left panels:} Time evolution of the equivalent width 
of the strongest WR lines (upper left) and WR/O star ratios including 
subtypes (lower left) at Z=0.004 for standard evolutionary models. 
{\em Right panels:} Models including massive close binaries for $f$=0.2.
Upper right: same as upper left.
Lower right: evolution of the relative \heii/\hbeta line intensities as
a function of the \hbeta\ equivalent width (solid line). The dashed line 
shows the contribution from single stars. The two ``epochs'' where WR
stars are formed from single stars and by the binary channel are well
separated in this diagram}
\label{fig_e004}
\end{figure}

\section{Probing WN and WC populations in starbursts}
We will illustrate the model predictions for a burst with a metallicity of $Z=0.004$, 
a typical value for the WR-galaxies analyzed by Vacca \& Conti 
(1992). [The entire set of results, which depend strongly on metallicity, 
will be discussed in Schaerer \& Vacca (1996).]
The left panels in Figure 1 present the results from our standard models 
(Salpeter IMF, instantaneous burst, single star evolution), while the right 
panels include massive close binary stars (cf.~Sect.~4).
The lower left Figure shows the relative WR and populations as a function 
of the age of the burst. The WR-rich phase lasts from $\sim$ 2.5 to 
5.5 Myr. WC stars evolving from the most massive stars 
dominate the WR population from about $3 - 4$ Myr, while
WNL stars are more numerous from $4 - 5.5$ Myr. Thus, the models predict
a WC-rich phase shortly after the first appearance of
WR stars. For an instantaneous burst the last period of the 
WR-rich phase is always dominated by WNL stars, as these objects represent
the descendents of the least massive stars which barely manage to peel
off their outer layers revealing the processed material resulting from H-burning.

The upper left panel shows the corresponding evolution of the equivalent 
widths of the most important WR lines.
The \heii\ and N~{\sc iv} 4640 emission predicted {\em before} the WR rich phase
is due to the (relatively large) contribution adopted for OIf stars.
Broad \heii\ 4686 emission usually dominates the optical spectrum except during 
the short ($\sim$ 1 Myr) WC-rich phase, during which C~{\sc iii/iv} 4650 dominates 
the broad classical ``WR bump'' and the presence of WC stars can be unambiguously 
deduced from the strong C~{\sc iv} 5808 feature. 
Although the predicted strength of N~{\sc iv} 4640 is relatively uncertain, its
is always lower than that of \heii\ 4686 except at solar or higher
metallicities. This is an immediate consequence of the abundance effect
pointed out by Smith \etal (1996).
As expected, C~{\sc iii} 5696 is very weak at $Z=0.004$; this feature is strong
only in late WC stars, which are not found at low metallicities.

\section{The frequency of WR-rich starbursts}
The predictions illustrated above can be used to determine the WR content
in individual starbursts and allow us to determine separately the WN and WC
populations. In addition to the study of individual objects, however,
a statistical analysis of a set of starburst galaxies also provides 
a test of the models, as recently stressed by Meynet (1995).
Although large samples adequate for such statistical 
studies are not yet available, we would like to point out briefly some 
interesting results from the low metallicity samples of Izotov \etal (1994, 1996a) 
and Pagel \etal (1992), which have been obtained as part of a systematic determination 
of the primordial He abundance. 
Since the major goal of these studies is to obtain as many low metallicity
objects as possible these samples are suited for statistical studies of
starbursts over a low, and clearly specified metallicity interval.

The Izotov sample contains 33 objects with $Z$ between $\sim 0.001$ 
and 0.004, of which 14 exhibit WR features, including 4 WC with signatures. 
Thus $\sim$ 40 \% of the objects show evidence of WR stars, and $\sim$ 30 
\% of those include WC stars.
Similar, or even larger, percentages of WR detections are found in the
Pagel \etal sample over a similar metallicity range.
Interestingly these numbers are fairly close to the percentages 
of starbursts containing WR stars 
predicted from evolutionary models (Meynet 1995)\footnote{The values 
for the high mass loss models at $Z=0.004$ in Table 1 of Meynet (1995) 
are erroneous. Furthermore the duration of the WC-rich phase given by
Meynet is overestimated. 
This accounts for the difference between our 
results in Fig. 1, and those in Fig. 3 of Meynet (1995).}.
The expected percentage is between $\sim$ 18 and 40 \% for $Z$ between 0.001 and 
0.004; the duration of WC-rich phase is predicted to be $\sim$ 1/3 of the WR phase
(see Fig.~\ref{fig_e004}).
An observational bias is introduced by the requirement that the
[O~{\sc iii}] 4363 line can be detected and reliably measured. This requirement
favours inclusion of objects with the youngest bursts and could therefore lead 
to an overestimate of the percentage of WR-rich objects
as compared to the definition used by Meynet (1995). 
This might be responsible for the apparent difference with
the model predictions at low $Z$ .

The approximate agreement between models and observations regarding 
the statistical number of WR-rich objects is very encouraging
although admittedly the present samples are fairly small.
In particular the detection of a significant fraction of WC stars at low
metallicities gives strong support to the adopted high-mass-loss 
evolutionary models. 
In this context it is also interesting to note that to date no
WR features have been detected in objects with metallicities below O/H 
$\le$ 7.7--7.8 (Pagel \etal 1992; Izotov \etal 1996), corresponding to an 
absolute metal abundance of $Z \le$ 0.0012--0.0015, or about $0.06 Z_\odot$.
Although no formal low metallicity cut-off for the presence of WR stars is expected
from evolutionary models, this observed limit seems to be in fair agreement
with the predicted sharp decrease in the duration of the WR phase 
between $Z=0.004$ and $0.001$ (cf.~Meynet 1995).

\section{The role of massive close binaries in young starbursts}
Recent studies have begun to explore the importance of the formation of WR 
stars in massive close binaries (MCB's) on massive star populations in 
starbursts (Cervi\~{n}o \& Mas-Hesse 1996; Vanbeveren \etal 1996).
Here we briefly discuss some basic considerations, which are useful to estimate 
those circumstances in which binary stars may be of relevance for the WR 
populations in starbursts.
An important property of binary models is that, because the high mass loss
rate prevents a large increase in the stellar radius, primaries with initial 
masses $M_1 >$ 40-50 \msun\ should, in general, avoid Roche lobe overflow 
(cf.~Vanbeveren 1995); for those stars that do experience Roche lobe overflow,
their evolution is nearly indistinguishable from that of single stars (Langer 1995). 
{\em Therefore, in instantaneous bursts with ages 
$\leq 5$ Myr the stellar population is unaltered by the formation of 
WR stars through the binary channel.}

{\em Do WR galaxies contain a significant population of WR stars formed 
through the binary channel ?}
The observed \hbeta\ equivalent width in the spectrum of an H II region exhibits 
a monotonic decrease with time can be used as a good indicator of the age of
a starburst (e.g., Leitherer \& Heckman 1995). Bursts with ages $\tau \ge 5$ 
Myr are predicted to have $W(\hbeta) <$ 60 \ang\ for $Z \sim 0.001$, while at larger 
metallicity the upper limit for $W(\hbeta)$ is even lower.
An inspection of the compilation of WR galaxies given Conti (1991) 
reveals that most objects have large \hbeta\ equivalent widths:
12 out of 37 objects have $W(\hbeta) <$ 60 \ang, and only 3 show
$W(\hbeta) <$ 30 \ang. In fact, because of various physical effects which serve to
artifically reduce the observed $W(\hbeta)$, these fractions are actually 
{\em upper} limits to the true number of WR galaxies with low equivalent widths.
Therefore, most WR galaxies experienced bursts of star formation
less than 5 Myr ago.
If star-formation has taken place on such a short timescale compared to the
lifetime of massive stars (``instantaneous burst'') roughly 70 to 90 \% 
of the burst populations in WR galaxies are too young to be affected
by WR formation through the binary channel and therefore they should
be well described by single star models.

{\em The link between population synthesis models and observable
quantities.}
In recent studies Cervi\~{n}o \& Mas-Hesse (1996) and Vanbeveren \etal 
(1996) have included massive close binaries (MCB's) in population synthesis models.
They find that
{\em (1)} the WR-rich phase of a starburst lasts much longer (up to 12-20 Myr)
when MCBs are taken into account,
{\em (2)} WR/O number ratios can be larger than those predicted by synthesis models
including only single stars, and
{\em (3)} even with a ``standard'' IMF the observed WNL/O ratios are well 
reproduced by their models (Vanbeveren et al.\ 1996).
These findings require some remarks.

As shown above, in the vast majority of the observed WR galaxies the
bursts are very young and therefore, in general, their WR population has probably not 
been formed through the binary channel. Older objects with a possibly large WR population
remain to be found; however most searches are biased against finding such objects.
Given the young age of the known WR galaxies, the ``observed'' WNL/O star ratios
of Vacca \& Conti (1992) cannot be compared to the large values obtained by 
Vanbeveren \etal (1996) in the ``binary rich'' WR phase.
Moreover, as shown by Schaerer (1996) the observed WR/O population can be
explained with single star models and a ``standard'' Salpeter IMF.

To allow for a direct comparison between synthesized stellar populations and
observations the relevant observable quantities (line fluxes, equivalent widths
etc.) need to be modeled (see Sects.~2 and 3).
Predictions from exploratory calculations which also include binary stars are 
shown in Fig.~\ref{fig_e004} (right panel).
The behaviour of the equivalent widths of the most important broad WR lines 
(upper right) nicely illustrates the prolonged WR phase.
The lower right panel shows that, compared to the flux in \hbeta, a relatively 
large flux in the broad \heii\ 4686 line can be obtained if binaries are included.
However, as mentioned before, such behaviour can be obtained only at ages 
$\tau >$ 5 Myr corresponding roughly to $W(\hbeta) <$ 30--60 \ang.

{\em Massive X-ray binaries as the origin of nebular HeII emission ?}
Based on the same age considerations we would like to mention several arguments
regarding the role of high-mass X-ray binaries (HMXRB) in the origin of 
{\em nebular} \heii\ emission in extragalactic \hii\ regions (see Garnett \etal 1991; 
Schaerer 1996). There are several lines of evidence that indicate that HMXRBs are 
{\em not} the source of the nebular \heii\ emission:
{\em (1)}
All the objects from the samples of Campbell \etal (1986) and Izotov \etal 
(1994, 1996ab) have large \hbeta\ equivalent widths, corresponding
to burst ages of less than $\sim$ 5 Myr. 
If nebular \heii\ emission is due to HMXRB these systems must have had 
primaries with very large masses ($M_1 \ge$ 40-50 \msun) necessary 
to form neutron star remnants. Such a scenario for the formation of HMXRB
seems to be very unlikely (van den Heuvel 1994).
The age argument was also put forward by Motch \etal (1994).
{\em (2)} Given the short duration of the X-ray emitting phase 
($\sim 5 \times 10^4$ yr, van den Heuvel 1994), it is very difficult to produce HMXRB 
in large numbers (e.g.~comparable to the number of equivalent O7 stars
in SBS 0335-052 according to Izotov \etal 1996b).
{\em (3)} It is not clear why the spatial distribution of nebular \heii\ should 
preferentially follow the continuum instead of the remaining emission lines 
as found by (Izotov \etal 1996b).
{\em (4)} Motch et al.\ (1994) find that the \heii\ emission and the X-ray 
emission are not spatially coincident, as would be expected if HMXRB are the source of
the \heii\ emission.

The above results render the MXRB hypothesis rather unlikely. 
In many objects WR stars appear to be a very likely source of the high energy photons
needed to ionize He II (Motch et al.\ 1994; Schaerer 1996) although peculiar O stars
close to the Eddington limit (Gabler \etal 1992) cannot be excluded.
We also note that out of the 38 objects 
from Campbell \etal and Izotov \etal (1994, 1996a) which have a definite measurement of 
\heii, only 7 objects are found at very low metallicities (O/H $<$ 7.72), 
for which WR features have never been detected.
 
%
%
{\small
\section*{Acknowledgements}
We thank Paul Crowther for providing us with observational data.
The work of DS is supported by a grant of the Swiss National Foundation 
of Scientific Research. Additional support from the Directors
Discretionary Research Fund of the STScI is also acknowledged.
WDV acknowledges support in the form of a fellowship from the 
Beatrice Watson Parrent Foundation.
}
%
%
\vspace*{-0.5cm}
 
\beginrefer
\vspace*{-0.5cm}
{\small

\refer Arnault Ph., Kunth D., \& Schild H., 1989, A\&A 224, 73

\refer Campbell A., Terlevich R., \& Melnick J., 1986, MNRAS 223, 811

\refer Cervi\~{n}o M., \& Mas Hesse J.M., 1994, A\&A 284, 749

\refer Cervi\~{n}o M.,  \& Mas Hesse J.M., 1996, 
ASP Conf. Series, Vol.~98, p.~174

\refer Conti P.S., 1991, ApJ 377, 115

\refer Contini T., Davoust E.,  \& Consid\`ere S., 1995, A\&A 303, 440

\refer de Loore C.,  \& Vanbeveren D., 1994, A\&A 292, 463


\refer Gabler R., Gabler A., Kudritzki R.P., M\'endez R.H., 1992, A\&A 265, 656

\refer Garnett D.R., Kennicutt R.C., Chu Y.-H.,  \& Skillman E.D., 1991, ApJ 373, 458

\refer Gonz\'alez-Delgado R.M., et al., 1994, ApJ 437, 239

\refer Heckman T., et al., 1996, ApJ, submitted

\refer Izotov Y.I., Thuan T.X.,  \& Lipovetsky V.A., 1994, ApJ 435, 647

\refer Izotov Y.I., Thuan T.X.,  \& Lipovetsky V.A., 1996a, ApJ, submitted

\refer Izotov Y.I., et al., 1996b, ApJ, submitted


\refer Kunth D.,  \& Joubert M., 1985, A\&A 142, 411 

\refer Kr\"uger H., Fritze-v. Alvensleben U., Fricke K.J., Loose H.-H.,
        1992, A\&A 259, L73

\refer Langer N., 1995, in ``Wolf-Rayet Stars: Binaries, Colliding Winds,
	Evolution'', IAU Symp.~163, Eds. K.A. van der Hucht, P.M. Williams,
	Kluwer, Dordrecht, p.~15 

\refer Leitherer, C. \& Heckman, T. M. 1995, ApJS 96, 9

\refer Meynet G., 1995, A\&A 298, 767

\refer Motch C., Pakull M. W., \& Pietsch W. 1994, in ``Violent Star Formation, 
From 30 Doradus to QSOs'', Ed. G. Tenorio-Tagle, Cambridge University Press, p. 208



\refer Pagel B.E.J., Simonson E.A., Terlevich R.J.,  \& Edmunds M.G., 1992, 
	MNRAS 255, 325 



\refer Schaerer D., 1996, ApJ 467, L17

\refer Schaerer D., de Koter A., Schmutz W., \& Maeder A., 
	1996a, A\&A 310, 837

\refer Schaerer D., de Koter A., Schmutz W., \& Maeder A., 
	1996b, A\&A 312, 475

\refer Schaerer D.,  \& Vacca W.D., 1996, ApJ, in preparation

\refer Schaller G., Schaerer D., Meynet G.,  \& Maeder A., 1992, A\&AS 96, 269

\refer Schmutz W., Leitherer C.,  \& Gruenwald R., 1992, PASP 104, 1164

\refer Smith L.F.,  \& Maeder A., 1991, A\&A 241, 77

\refer Smith L.F., Shara M.M.,  \& Moffat A.F.J., 1990a, ApJ 348, 471 

\refer Smith L.F., Shara M.M.,  \& Moffat A.F.J., 1990b, ApJ 358, 229 

\refer Smith L.F., Shara M.M.,  \& Moffat A.F.J., 1996, MNRAS 281, 163

\refer Vacca W.D.,  \& Conti P.S., 1992, ApJ 401, 543

\refer Vanbeveren D., 1995, A\&A 294, 107

\refer Vanbeveren D., Van Bever J.,  \& De Donder E., 1996, A\&A, in press

\refer van den Heuvel E.P.J., 1994, in ``Interacting Binaries'', Saas-Fee Advanced
Course 22, Eds. H. Nussbaumer, A. Orr, Springer, p.~263


}
\endrefer 
          
\end{document}